\documentclass[11pt]{article}
 \usepackage{amssymb}
 \usepackage{amsthm}
 \usepackage{amssymb,amsmath,exscale}
\usepackage{color}
 \newcommand{\zProof}{{\noindent\bf\underbar{Proof}.}\ }

\usepackage{amsmath}
\usepackage{amsfonts}
\usepackage{amssymb}
\newcommand{\zdia}{~~\rule{1mm}{2mm}\par\medskip}
   
\newtheorem{thm}{Theorem}

\newtheorem{lem}[thm]{Lemma}
\newcommand{\ZD}{\;\mbox{\rm d}}

\newcommand{\ty}{\infty}
\newcommand{\di}{\displaystyle}
\newcommand{\va}{\varphi}
\newcommand{\si}{\sigma}
\newcommand{\ga}{\gamma}

\newcommand{\te}{\theta}

\newcommand{\noa}{\noalign{\medskip}}
\newcommand{\la}{\lambda}

\newcommand{\qu}{\quad}
\newcommand{\fo}{\forall}

\newcommand{\pa}{\partial}
\newcommand{\ti}{\times}

\numberwithin{equation}{section}

\begin{document}

 {\bf\Large Approximate controllability and lack of controllability to zero of the heat equation with memory\footnote{ The research of the first author is partially supported by Romanian CNCS Grant PN-II-ID-PCE-2011-3-0211. The research of the second author fits the  plans of INDAM-CNR and of the project ``Groupement de Recherche en Contr\^ole des EDP entre la France
et l'Italie (CONEDP)''.}  }

\bigskip
 
 \begin{center}
 Andrei Halanay\footnote{ 
 Department of Mathematics and Informatics, University Politehnica of Bucharest, 
  313 Splaiul Independentei, 060042 Bucharest, Romania,  halanay@mathem.pub.ro  }  
  
 Luciano Pandolfi\footnote{ Dipartimento di Scienze Matematiche ``G. L. Lagrange'', Politecnico di Torino, \\ Corso Duca degli Abruzzi 24, 10129 Torino, Italy,  luciano.pandolfi@polito.it} 

 \end{center}

\noindent
{\bf\underline{Abstract}}

\noindent
In this paper we consider the heat equation with memory in a bounded region $\Omega \subset\mathbb{R}^d$, $d\geq 1$, in the case that the propagation speed of the signal is infinite (i.e. the Colemann-Gurtin model). The memory kernel is of class $C^1$. We  examine its controllability properties both under the action of boundary controls or when the controls are distributed in a subregion of $\Omega$. We prove approximate controllability of the system and, in contrast with this, we prove   the existence of initial conditions which cannot be steered to hit  the target $0$ in a certain time $T$, of course when the memory kernel is not identically zero. In both the cases we derive our results from well known properties of the heat equation.

\bigskip

\noindent
{\bf\underline{{keyword}}}
Heat equation with memory, approximate controllability, controllability to zero, lack of controllability



\section{Introduction}
\label{}

This paper is concerned with controllability properties of the heat equation with memory (\cite{CG,JP})
\begin{equation}\label{1.1}
\theta'=\frac{\partial \theta}{\partial t} (t,x) = a\theta (t,x)+\Delta  \theta (t,x) + \int^t_0 M(t-s) \Delta  \theta (s,x) ds +F(t,x)
\end{equation}
where $x\in \Omega\subseteq \mathbb{R}^d$ is a bounded region with $C^2$ boundary (which lays on one side of its boundary)  $t \in [0,T]$,   $a$ is a real constant whose role will be specified at the end of this section and
  $\Delta = \Delta_x$ is the laplacian in the space variable $x$.

We assume that the kernel $M(t)$ is of class $C^1$ (in fact, $H^1$ would be enough).

We associate an initial condition  to Eq. (\ref{1.1}):
\begin{equation}\label{1.2}
\theta (0,x) = \xi (x), \quad x\in \Omega
\end{equation}
and a boundary condition defined on an open set $\Gamma \subset \partial \Omega$ by
\begin{equation}\label{1.3}
\theta (t,x) = \left\{\begin{array}{cl} f(t,x), & x \in \Gamma, \; t \in [0,T] \\ \noalign{\medskip} 0, & x \in \partial \Omega \setminus \Gamma, \; t \in [0,T] \end{array} \right.
\end{equation}

In the following applications the affine term $F(t,x)$ has the form
\begin{equation}\label{eq:formaControllodistrib}
F( t,x)=u(t,x)\chi_{\omega}(x), \quad \omega\subset\Omega\,,
\end{equation}
where  $ \chi_{\omega}(x)$ is the characteristic function of the region $ \omega $ so that
 $u$ is a distributed control which acts in the subregion $\omega\subseteq\Omega$ 
 while $f$ is a control function acting on   $\Gamma$.

When studying controllability, we shall assume that either $f=0$, and $u $ is the \emph{active control,} or $u=0$ and $f $ is the \emph{active control.}

Note that dependence of the functions on the variables $t$ and $x$ is indicated only when needed for clarity, and in general we shall  write $\theta$ or $\theta(t)$ instead of $\theta(t,x)$.

Eq. (\ref{1.1}) is called the Coleman-Gurtin (or Jeffrey) model for thermodynamical systems with memory, see~\cite{JP}.

The solutions of problem (\ref{1.1})-(\ref{1.3}) are defined in Section 2, where we prove also the following result:

\noindent
\begin{thm}
\label{Theorem1.1.} {  Suppose $M \in C^1 (0,\infty)$.  For every $T > 0$,   $f \in L^2 ((0,T)$, $L^2 (\Gamma))$, $F\in L^2((0,T)$; $L^2(\Omega))$ and for every initial condition $\xi \in L^2 (\Omega)$ we have}:

\begin{enumerate}
\item\label{teo1:item1}  {  there exists a  unique solution $\theta  \in L^2 ((0,T), L^2 (\Omega))$ of} (\ref{1.1})-(\ref{1.3}). The transformation $(\xi,F,f)\mapsto \theta$ is linear and continuous in the specified spaces.

\item\label{teo1:item2} {  If there exists $\varepsilon>0$ such that   $ f(t,x)=0$ for $ t\in (0,\epsilon ) $ or $t\in (T-\varepsilon,T)$ then}   $ \theta\in C([0,\epsilon );L^2(\Omega))$ or $ \theta\in C((T-\varepsilon,T];L^2(\Omega))$.
\end{enumerate}

\end{thm}
The previous result and a counterexample in \cite{LL} show that the evaluation of $\theta$ at a fixed time $T$, which is a crucial ingredient in controllability, is meaningless if the boundary control is merely of class $L^2((0,T);L^2(\Gamma))$.  For this reason we define:
 
{\bf {\underline{Definition}.} } A boundary control $f\in L^2((0,T);L^2(\Gamma))$ is \emph{admissible} when there exists $\varepsilon >0$ such that  $t\mapsto \theta(t,x)\in C((T-\varepsilon,T],L^2(\Omega))$ (the number $\varepsilon$ depends on the control $f$).

\medskip

As stated in Theorem~\ref{Theorem1.1.},  sufficient condition for admissibility is that $f=0$ in $(T-\varepsilon,T)$, a condition usually imposed when giving sufficient conditions of controllability, see~\cite{ZUAZUAcontroll}.

Now we define:

{\bf {\underline{Definition}.}}
 Let $F(t,x)=u(t,x)\chi_{\omega}(x)$.

\begin{enumerate}
\item
A target $ \eta\in L^2(\Omega) $ is \emph{reachable} in time $ T $  under the action of the \emph{distributed control}  when there exists $ u\in L^2([0,T];L^2(\Omega)) $ such that $ \theta(T  )=\eta  $ (we assume $ \xi=0 $ and $ f=0 $); it is \emph{ reachable} in time $ T $ under the action of the \emph{boundary control  } when there exists an admissible boundary control $ f\in L^2([0,T];L^2(\Gamma))$ such that  $\theta(T )=\eta $ (here we assume $\xi=0 $, $u=0$).

The set of the reachable targets at time $T$ is the \emph{reachable set} (either under the distributed or boundary control) at time $T$. It is denoted $\mathcal {R}_T$.

\item the system is \emph{approximately controllable} (under the distributed or boundary control) when the corresponding reachable set is dense in $ L^2(\Omega) $.

\item
System (\ref{1.1}) is \emph{controllable to zero} in time $ T $ when for every $ \xi\in L^2(\Omega) $ there exists a  control $ u $   such that, with $ f=0 $, we have $  \theta(T)=0 $; alternatively, there
exists an admissible control $ f $ such that, with $ u=0 $, we have $\theta(T)=0 $.
\end{enumerate}

\medskip

{\bf {\underline{Remark}.}}
\label{Remark1.} We defined controllability with the initial condition equal zero and the control which is not active equal zero too. It is well known that for linear systems approximate or null controllability with \emph{any fixed} initial condition and/or \emph{fixed} non active control holds if and only if it holds with these elements put equal zero.

The main results of this paper are:

\noindent
\begin{thm}
\label{Theorem1.4.} {  If $M \in C^1 (0,\infty)$, approximate controllability holds for} (\ref{1.1}) {  both under the action of the distributed or the boundary control}.
\end{thm}

\medskip

A second result is that we prove lack of controllability to zero.  Of course here it is crucial that the memory kernel be nonzero. This is expressed in terms of the
\emph{resolvent kernel } of $M(t)$.
We first recall the following property of the Volterra integral equation in $\mathbb{R}$ (see \cite{GL}):
$$
y(t) + \int^t_0 M(t-s) y(s) ds =g(t):
$$
the unique solution that exists for $g \in L^2_{{\rm loc}} ([0,\infty))$   is given by
\begin{equation}\label{2.1}
y(t) = g(t) - \int^t_0 R(t-s) g(s) ds
\end{equation}
where $R$, called the resolvent kernel of $M$, solves

\begin{equation}\label{2.2}
R(t) = M(t) - \int^t_0 M(t-s) R(s) ds.
\end{equation}
Furthermore, the transformation $g\to y$ is linear, continuous and continuously invertible from $L^2(0,T)$ to itself for every $T>0$.

The following result holds true:

\begin{thm}
\label{Theorem1.5.} { Let $R$ be the resolvent kernel of $M \in C^1 (0,\infty)  $ and let $T>0$  be such that $R(T) \ne 0$.
Then we have}:

\begin{itemize}
\item  {  let $\omega\not= \Omega$. Then there exist an initial data $\xi \in L^2 (\Omega)$ that cannot be controlled to zero at time $T$ by an interior control};

\item {  there exist an initial data $\xi \in L^2 (\Omega)$ that cannot be controlled to zero at time $T$ by an admissible boundary control}.
\end{itemize}
\end{thm}

A word of explanation is needed to understand properly the sense of this theorem: when proving controllability to zero of the (memoryless) heat equation, it is assumed that the boundary control is (for example) zero in a first interval $ (0,\epsilon ) $
so to have continuity of the solution on this interval. 
In the study  of   controllability to zero under boundary control we prove even a stronger result: 
we shall give a formula for the solution $ \theta(t) $ of~(\ref{1.1}), which depends on $ \xi $. We shall prove that
there exist elements $ \xi\in L^2(\Omega )  $ such that $ \theta(T)=0  $ cannot be achieved by any admissible control, not even when when $ \theta(t) $  is not continuous for $ t $ close to zero.

{\bf {\underline{Remark}.}}(The role of the constant $a$)
\label{Remark2}The constant $a$ has no role in the study of controllability, since the transformation $w(t,x)=e^{-\gamma t}\theta(t,x)$ transforms Eq. (\ref{1.1}) to
$$
w'  =(a-\gamma)w +\Delta w   + \int^t_0 \left [ M_\gamma(t-s)\right ] \Delta  w (s ) ds +e^{-\gamma t}F 
$$
$$
M_\gamma(t)=e^{-\gamma t}M(t)
$$
with the same initial condition and boundary control $e^{-\gamma t}f(x,t)$.

The important property is that
\[
M_\gamma(0)=M(0)\,.
\]
Hence, \emph{we can change at will the value of $a$, without changing the value of the kernel at $t=0$.} We use this observation since the following arguments are slightly simplified if we use the previous transformation to assign a special value to the constant $a$:
\begin{equation}\label{eq:DefiCONSTANTa}
a=-M(0)\;.
\end{equation}

\subsection{References and the goals of this paper}

Heat equations with memory of the Gurtin-Pipkin type (see \cite{GP}), i.e.
\[
\te'=\int_0^t M(t-s)\Delta \te(s) ds
 \]
has been widely studied, (see \cite{AP,FYZ,KK,LPS,PP,PP1,PP2,PP3}). Instead, the Colemann-Gurtin model received far less attention. It seems that Theorem~\ref{Theorem1.4.} has been first proved in \cite{BI} (see also~\cite{IP,XZ}) when the kernel $ M(t) $ is of the form
\[
 M(t)=\sum _{k=1}^m \int_{I_k} b_k(s) e^{-st} ds  +\sum _{j=1}^l a_j e^{-\la_j t}
\]
where $a_j\geq 0$, $\la_j\geq 0$ and the functions $b_k(t)$ are integrable and nonnegative on the intervals $I_k$.  We extend this result to any $C^1$ kernel.

Theorem~\ref{Theorem1.5.} (and boundary control) has been proved in \cite{GI} when $ M(t)\equiv 1 $ (see also \cite{Rosier,XZ})     and in \cite{HP} for a certain   kernel  which satisfies  the restrictions imposed by thermodynamics, but dimension $ d=1 $, i.e. when $ \Omega $ is an interval. Then, the result in \cite{GI, HP} has been extended to every $ C^1 $ kernel (see \cite{HP1}), still when $ \Omega $ is an interval.

\emph{A first goal} of the present paper is an extension of the previous results  to domains $\Omega$ in $ \mathbb{R}^d $, when $ d>1 $ and any (smooth) kernel.

The cited references prove approximate controllability or lack of   controllability to zero using delicate estimates on     certain   sequences of exponentials, or their biorthogonal sequences, which extend the estimates 
first given in~\cite{FR}. 

\emph{Our second goal}  is to show that this step can be skipped since
we derive our results directly as special instances of known properties of the memoryless heat equation. 

\section{Definition of the solutions and the proof of Theorem~\ref{Theorem1.1.}}

Let $A$ be the operator
\begin{equation}\label{1.4}
A :\hbox{dom} \: A \mapsto L^2(\Omega), \qquad
\hbox{dom}\: A = H^2 (\Omega) \cap H^1_0 (\Omega)\, ,\qquad Au = \Delta u\;.
\end{equation}
It is known that the operator $A$ is selfadjoint, with compact resolvent and bounded inverse $A^{-1}$; it  generates a holomorphic semigroup $e^{At}$ (see \cite{BPD, TW}). Closely related to $A$ is the Dirichlet operator $D$: $v = Df$ where $v$ solves
\begin{equation}\label{1.5}
\Delta v = 0 \; \hbox{in} \; \Omega, \quad   \left\{\begin{array}{ll}v(x)= f(x), & x \in \Gamma,  \\   v(x)=0, & x \in \partial \Omega \setminus \Gamma\,. \end{array} \right.
\end{equation}
For details about $D$, see~\cite{BPD, TW}).

In order to define the solutions of Eq. (\ref{1.1}) we use a formal  computations (involving the  MacCamy trick) and we reduce the equation to a Volterra integral equation in $L^2(\Omega)$. The solutions of this Volterra integral equations are by definition the solutions of (\ref{1.1}).

When (\ref{2.1}) is used in (\ref{1.1}) with $y(t) = \Delta  \theta (t,x)$ and $g=\theta'-F-a\theta$, we formally get
\begin{equation}\label{2.3}
\begin{array}{l}
\displaystyle
\di\frac{\partial\theta}{\partial t} (t,x) = \Delta  \te (t,x)+a \te (t,x) +\\
\displaystyle + \int^t_0 R(t-s)  \frac{\partial\theta}{\partial s} (s,x) ds  - a \displaystyle\int_0^t R(t-s)\te(x,s) ds + G(t)\,,  \\
\displaystyle  G(t)=F(t)- \di\int^t_0 R(t-s)F(s)\ZD s\,.
\end{array}
\end{equation}
This formal computation is known as \emph{MacCamy trick.}

Note that the transformation $F\mapsto G$ is linear, continuous and continuously invertible in $L^2(0,T;L^2(\Omega))$ for every $T>0$, see \cite{LR}.
Introduce
\begin{equation}\label{2.4}
L(t) = R'(t)-aR(t)=R'(t)+M(0)R(t)
\end{equation}
(by (\ref{2.2}), $R$ is $C^1$ and $R(0)=M(0)$). An integration by parts in (\ref{2.3}) gives, using (\ref{1.2}) and (\ref{eq:DefiCONSTANTa}), i.e.    $ a+R(0)=a+M(0)=0 $:
\begin{equation}\label{2.5}
\di\frac{\partial\theta}{\partial t} (t,x) = \Delta  \te (t,x)   - R(t) \xi (x) + \displaystyle\int^t_0 L (t-s) \te (s,x) ds +G(t,x)
\end{equation}
A solution of (\ref{2.5}) will be, by definition, a solution  of (\ref{1.1})-(\ref{1.3}).

Now we recall the following result concerning the (memoryless) heat equation (see for example \cite[p.~7]{LT})

\begin{thm}
  \label{t2.1}  {  Suppose $f \in L^2 ((0,T), L^2 (\Gamma))$, $\xi \in L^2 (\Omega)$, $g \in L^2 ((0,T) \ti \Omega)$. The solution of the following mixed problem
\begin{equation}\label{2.6}
\di\frac{\partial w}{\partial t} (t,x) = \Delta  w(t,x) + g(t,x)\quad
\left\{\begin{array}{l}
 w (0,x) = \xi (x), \; x \in \Omega \\   w (t,x) = f(t,x), \; t \in [0,T), \; x \in \Gamma\\
w (t,x) = 0, \; t \in [0,T), \; x \in\partial\Omega\setminus \Gamma 
 \end{array}\right.
\end{equation}
is unique in $L^2 ((0, T), L^2 (\Omega))$  and is given by}
\begin{equation}\label{2.7}
w (t )  = e^{At} \xi    + \di\int^t_0 e^{A(t-s)} g(s  ) ds   
 - A \di\int^t_0 e^{A(t-s)} (D  f)(s 
 ) ds \,.
\end{equation}

{  The transformation $(\xi, f,g)\mapsto w$ is  continuous in the specified spaces. Furthermore we have}:
\begin{itemize}
\item {  if $\xi$, $g$ and $f$ are of class $C^\infty$ with compact support respectively in $\Omega$, $(0,T)\times\Omega$ and $(0,T)\times\Gamma$ then $w(t,x)$ has continuous first derivative in $t$ and second derivatives in the space variable}.
\item {  if $f(t)=0$ on $ (0,\epsilon) $ or on $(T-\varepsilon,T)$ then $t\mapsto w(t,x)$  } is an $ L^2(\Omega) $-valued   function which is continuous on $[0,\epsilon)$ or on $(T-\epsilon,T]$.
\end{itemize}
\end{thm}
Now we use a last formal step in Eq. (\ref{2.5}): {  we apply formula} (\ref{2.7}) {  with}
\begin{equation}\label{2.8}
g(t ) = \int^t_0 L(t-s)\te (s ) ds - R(t) \xi   +G(t )\,.
\end{equation}
We get the following integral equation for $\te$
\begin{equation}\label{2.9}
\begin{array}{c}
\theta (t ) - \displaystyle\int^t_0 e^{A(t-s)} \left[  \displaystyle\int^s_0 L(s-r)\te (r ) dr \right] ds = \\ \noa = e^{At} \xi   - \displaystyle\int^t_0 e^{A(t-s)} R(s) \xi   ds
+ \\ + \displaystyle\int_0^t e^{A(t-s) }  G(s )  ds
 - A \displaystyle\int^t_0 e^{A(t-s)}  D  f (s ) ds\,. \end{array}
\end{equation}

The properties of the Volterra integral equations in Hilbert spaces  (see for example \cite{LR})
 show the existence of a unique solution $ \theta(t)\in L^2(0,T;L^2(\Omega)) $ of~(\ref{2.9}), which depends continuously on $ \xi $, $ f $ and $ F $ as specified in Item~\ref{teo1:item1} of Theorem~\ref{Theorem1.1.}.

{\bf {\underline{Definition}.}}
The unique solutions of~(\ref{1.1})-(\ref{1.3}) is by definition the unique solution of the Volterra integral equation (\ref{2.9}) in the Hilbert space $L^2(\Omega)$.

\medskip

In conclusion, Theorem~\ref{Theorem1.1.}   follows from  Theorem~\ref{t2.1}  and the properties of the Volterra integral equations in Hilbert spaces. Note in particular that the memory term in (\ref{2.9}), i.e. the integral on the left side, is a continuous function of time so that we have continuity of $\theta(t)  $ when the last integral on the right hand side is continuous, in particular on $ [0,\epsilon) $ or on $ (T-\epsilon,T] $ when $ f(t) $ is constant (in particular, equal zero) on these intervals.

Finally we state the following regularity result for the solutions of the  heat equation (\ref{2.6}), and which are inherited by the solutions of (\ref{1.1})-(\ref{1.3}):

\begin{lem}
\label{Lemma2.2.} { Let $T>0$ and let $\xi(x)$, $F(t,x)$  and $f(t,x)$  be of class $C^\infty$ and with compact support respectively in $\Omega$, $(0,T)\times\Omega$ and $(0,T)\times\partial\Omega$. Then the solutions $w (t,x)$
of (\ref{2.6})  have the following regularity property: for $y(t) = w(t)-Df(t) $, one has} $y\in C([0,T], {\rm dom}\, A)\cap C^1([0,T];L^2(\Omega))$.

This property is inherited by the solutions of (\ref{1.1})-(\ref{1.3}) in the following sense:

\begin{equation}
\label{eq:reGOLARiANTE}
y=\theta -Df\in C([0,T],{\rm dom}\, A)\cap C^1([0,T];L^2(\Omega))
\end{equation}
where now $\theta (t)$ is given by (\ref{2.9}).
\end{lem}
 \zProof
 We prove the property for the solution $ \theta(t) $ in~(\ref{2.9}). The statement for $ w(t) $ is the special case when $M(t)=0$, hence $ R(t)= L(t)=0 $.
The last integral in~(\ref{2.9}) (i.e. in~(\ref{2.7})) can be integrated by parts and we get
\begin{eqnarray*}
&&\theta(t)-Df(t)-\int_0^t e^{As}\int_0^{t-s}L(r) \theta(t-s-r) dr\, ds =\\
&&= e^{At} \xi  -  \int_0^t  e^{As} R(t-s)\xi ds+\int_0^t  e^{As} G(t-s) ds-\int_0^t  e^{At} Df'(t-s)ds \,.
\end{eqnarray*}
 So, $y(t)=\theta(t)-Df(t)$ solves
 \begin{eqnarray*}
&&y(t)-\int_0^t  e^{As}\int_0^{t-s}L(r) y(t-s-r) dr\, ds=\\
&& = e^{At} \xi  - \int_0^t  e^{A(t-s)} R( s)\xi ds 
  +\int_0^t  e^{As} G(t-s) ds-\int_0^t  e^{At} Df'(t-s)ds+\\
&&+\int_0^t  e^{As}\int_0^{t-s}L(r) Df(t-s-r) dr\, ds 
\end{eqnarray*}
and either $y'(t)$ or $Ay(t)$ are the solution of, respectively,
 \begin{eqnarray*}
&&y'(t)-\int_0^t  e^{As}\int_0^{t-s}L(r) y'(t-s-r) dr\, ds-\int_0^t  e^{As} L(t-s)\xi ds =\\
&& = Ae^{At} \xi  -  \int_0^t  e^{A(t-s)} R(s)A\xi ds -R(t)\xi+\int_0^t  e^{As} G'(t-s) ds- \\
&&  -\int_0^t  e^{At} Df''(t-s)ds+\int_0^t  e^{As}\int_0^{t-s}L(r) Df'(t-s-r) dr\, ds 
\end{eqnarray*}
or
  \begin{eqnarray*}
&&\left [ Ay(t)\right ] -\int_0^t  e^{As}\int_0^{t-s}L(r)  \left [ Ay(t-s-r) \right ] dr\, ds  =\\
&& = e^{At} A\xi  -  \int_0^t  e^{As} R(t-s)A\xi ds  +\\
&& +\int_0^t  \left [Ae^{As}\right ] \left (G(t-s) - Df'(t-s) + \int_0^{t-s}L(r) Df(t-s-r) dr\,\right) ds \,.
\end{eqnarray*}
The last integral  is an $L^2(\Omega)$-continuous function  since it can be integrated by parts.

These equalities show that $y'\in C([0,T];L^2(\Omega))$ and $Ay\in C([0,T];L^2(\Omega))$, as wanted.
 \zdia

This result can be used   to justify our definition of the solutions of (\ref{1.1})-(\ref{1.3}) since, when combined with continuous dependence on $ \xi $, $ f $ and $ F $,
shows that the solutions of (\ref{1.1})-(\ref{1.3})   defined by (\ref{2.9})  are limits of smooth solutions. Furthermore, it  justifies  the results of the  following computations, which are correct for ``smooth''  data, i.e. when the assumptions in Lemma 2.2 hold, and then extended by continuity to every solution.

\subsection{Projection on the eigenfunctions}
It is known (see e.g. \cite{TW}) that $L^2(\Omega)$ has an orthonormal basis $\{\va_n\} $ of eigenvectors of the operator $A$. Let $-\la_n^2$ be the eigenvalue of $\va_n$. Then:
\begin{enumerate}
\item at most a finite number of eigenvectors have the same eigenvalue;
\item we have $-\la_n^2<0$ for every $n$;
\item it is possible to order the elements of the basis $\{\va_n\} $ so that the sequence $\{\la_n^2\}$ is increasing.
\end{enumerate}

\begin{lem}\label{lemma:sulDomiA} The definition of the operator $A$ shows that:
\begin{enumerate}
\item the eigenvectors $\phi_n$ of $A$ satisfy 
\[
A\va_n=\Delta \va_n = -\la^2_n \va_n\,,\ \va_n\in H^2(\Omega)\cap H^1_0(\Omega)
\]
so that the trace $\ga_0\va _n$ on $\partial \Omega$ is well defined and it is zero and also $\ga_1\va _n$,  the exterior normal derivative on $\partial\Omega$, is well defined  and
\[
\ga_1\va _n=  \di\frac{\partial}{\partial \nu}\va _n \in L^2(\partial\Omega) \,.
\]
\item\label{item2:lemma:sulDomiA}
An element $\xi\in L^2(\Omega)$ belongs to ${\rm dom}\, A$ if and only if 
\[
\xi(x)=\sum _{n=1}^{\infty}\frac{c_n}{\lambda_n^2}\phi_n(x)\,,\qquad \{c_n\}\in l^2\,.
\]

\end{enumerate}
\end{lem}
The first  property is used to justify Green formula
 $$
\begin{array}{l}
\di\int_{\Omega}  (\Delta  \te (t,x) \va_n (x) dx - \te (t,x) \Delta \va_n (x)) dx = \\ \noa = \di\int_{\pa \Omega} \left[\di\frac{\pa}{\pa \nu} \te (t,x) \va_n (x) - \te (t,x)  \di\frac{\pa \va_n}{\pa \nu} (x) \right] d\si =   - \di\int_{\Gamma} f(t,x) (\gamma_1 \va_n) d \si \end{array}
$$
and the following equality   (see \cite[Prop. 10.6.1]{TW}):
\begin{equation}\label{eq:daTW}
\int_\Omega \va_n(x)( Df )(x) dx=-\frac{1}{\la_n^2}\int_\Gamma \left (\ga_1\va_n(x)\right )f(x) d\sigma\,.
\end{equation}
We represent;
\begin{equation}
\label{eq:serie ditheta}
\theta(t,x)=\sum _{n=1}^{\infty} \phi_n(x)\theta_n(t)\,,\qquad \te_n(t)=\int_\Omega \te(t,x)\va _n(x)  dx.
\end{equation}
Let
 \begin{align*}
    & \xi_n=\int_\Omega  \xi (x)\va _n(x)\ZD x\, \quad    F_n(t)=\int_\Omega F(t,x) \va _n(x) dx\,,
\\
& G_n(t)=F_n(t)-\int_0^t R(t-s)F_n(s) ds \,.
 \end{align*}

Using (\ref{eq:daTW}) in (\ref{2.9}) we get:
\begin{equation}
\label{eq:proezioneDItheta}
\begin{array}{l}
\displaystyle
\te_n(t)-\int_0^{t}\left [\int_0^{t-s} L(t-s-r)e^{-\la_n^2 r} dr\right ]\te_n(s) ds=\\
\displaystyle= \left [e^{-\la_n^2 t}-\int_0^t e^{-\la_n^2(t-s)}R(s) ds\right ]\xi_n+\\
\displaystyle +\int_0^t e^{-\la_n^2(t-s)} G_n(s) ds
-
\int_0^t e^{-\lambda_n^2(t-s)}\left [
\int_\Gamma f(s)\left (\gamma_1\phi_n \right )d\sigma
\right ]
ds\,.
 \end{array}
\end{equation}

For the sake of concision, we shall use the notation
\begin{equation}\label{3.4}
f_n (t) = \int_\Gamma (\ga_1 \va_n)   f(t,x) d\si\,.
\end{equation}

Equations (\ref{eq:proezioneDItheta}) are Volterra integral equations for $\te_n(t)$, with kernels
 \[
 Z_n(t)=-\int_0^{t} L(t-s) e^{-\la_n^2 s} ds
 \]

 Let $H_n(t)$ be the resolvent kernel of $Z_n(t)$:
  \[
H_n (t) = Z_n (t) - \int^t_0 Z_n (t-s) H_n (s) ds \,.
 \]
 Then we have:
\begin{align*}
 & \te_n (t)
=  e^{-\lambda ^2_n t}  \xi_n-\left [
 \int_0^{t} e^{-\lambda _n^2(t-s)}R(s)ds-\int_0^t H_n(t-s)e^{-\la_n^2 s} ds + \right.    \\
 &
\left.
  +\int_0^{t} R(s) \left (\int_0^{t-s} H_n(\tau) e^{-\lambda _n^2(t-s-\tau)} d\tau\right ) ds
\right ]\xi_n\\
&+\int^t_0\left [ e^{-\lambda ^2_n (t-s)}-\int_0^{t-s} H_n(\tau) e^{-\lambda _n^2(t-s-\tau)}d\tau    \right ]\left (G_n(s)-f_n(s)\right )ds\,.
\end{align*}
 The following result holds:

\begin{thm}
\label{Theorem2.3.} {  There exists a continuous function $J(t,s)$, which does not depend on $n$, such that}
\begin{equation}
\label{Eq:formadiHn}
\int^t_0 H_n (t-\tau) e^{-\lambda ^2_n \tau} d\tau=\int^t_0 J(t,\tau) e^{-\lambda ^2_n \tau} d\tau \;.
\end{equation}
{ Furthermore the following estimates hold for $t\in [0,T]$ (any $T>0$)}:
$$
|H_n (t)| \le \frac{M_T}{\lambda ^2_n},
$$
\begin{equation}\label{eq:STIMEdiHn}
 \left| \int^T_0 H_n (T-\tau) \left(e^{-\lambda ^2_n \tau} - \int^\tau_0 e^{-\lambda ^2_n (\tau -s)} R(s) ds \right) d\tau \right| \le \frac{M_T}{\lambda ^4_n}.
\end{equation}
\end{thm}
The proof is in \cite{HP1}. For completeness we report in the appendix the proof of the interesting equality (\ref{Eq:formadiHn}). Using this equality we can represent:
\begin{align}
\nonumber  & \te_n (t)
=  e^{-\lambda ^2_n t}  \xi_n-\left [
 \int_0^{t} e^{-\lambda _n^2(t-s)}R(s)ds-\int_0^t H_n(t-s)e^{-\la_n^2 s} ds + \right.     \\
\nonumber &
\left.
  +\int_0^{t} R(s) \left (\int_0^{t-s} H_n(\tau) e^{-\lambda _n^2(t-s-\tau)} d\tau\right ) ds
\right ]\xi_n\\
\label{FormaFINALthetaN}&+\int^t_0\left [ e^{-\lambda ^2_n (t-s)}-\int_0^{t-s} J(t-s,\tau)e^{-\la_n^2\tau }d\tau    \right ]\left (G_n(s)-f_n(s)\right )ds\,.
\end{align}

We are now in position to study the control properties of system (\ref{1.1}).

\section{Approximate controllability}

 We recall that when $M=0$, i.e. for the memoryless heat equation, approximate controllability holds both when the control acts on an open part of $\partial\Omega$ and when it is distributed in an open region $\omega\subseteq\Omega$. We prove that these controllability properties are inherited by the system with memory.

 As we noted, in the study of approximate controllability we can assume $\xi=0$ and either $f=0$ and $F(x,t)=\chi_\omega(x) u(t,x)$ or $F=0$ and $f$ the active control.

When $F(t,x)$ has the form~(\ref{eq:formaControllodistrib}) then
\[
G(t,x)=\chi_\omega(x)\tilde u(t,x)\,,\quad \tilde u(t,x)=u(t,x)-\int_0^t R(t-s)u(s,x) ds
\]
and $\tilde u\in L^2(0,T;L^2(\Omega))$ is arbitrary. Hence we can study controllability in terms of $G(t,x)=\chi_\omega(x)\tilde u(t,x)$.

Let $T>0$ be fixed. The set of the vectors $\theta(T)$ (reachable either under the distributed or the boundary control) is dense in $L^2(\Omega)$ when the sequences of their Fourier coefficients $\{\theta_n(T)\}$ are dense in $l^2$.

We consider first the case of the distributed control, i.e. the case $f=0$.
We must study the sequences whose elements are
 \begin{align*}
 \theta_n(T)&=
 \int_0^T e^{-\la_n^2 s}G_n(T-s) ds-\int_0^T \left [\int_0^{T-s}J(T-s,\tau)e^{-\la_n^2 \tau} d\tau\right ] G_n(s) ds=\\
   &= \int_0^T e^{-\la_n^2 (T-r)} \left \{G_n(r)-\int_0^r J(T-s,T-r) G_n(s) ds \right \} dr=\\
   &=\int_\Omega \phi_n(x)\chi_\omega(x)\int_0^T e^{-\la_n^2(T-r)}\left [u(r,x)-\int_0^r R(r-s)u(s,x) ds\right ]dr \, dx=\\
   &=\int_\Omega \va _n(x)\chi_\omega(x)\left \{\int_0^T
   e^{-\la_n^2(T-r)}\tilde u(r,x)          dr
   \right \} dx
 \end{align*}

 This formula holds also if $M(t)\equiv 0$  and in this case we have also $R(t)=0$ and $J(t,s)=0$.

 Now let
 \begin{equation}\label{eq:defiTARGET}
 \eta(x)=\sum _{n=1}^{\infty}\va _n(x) \eta_n\,,\qquad \{\eta_n\}\in l^2
\end{equation}
  be reachable for the memoryless heat equation so that there exists a control $\tilde u(t,x)\in L^2(0,T;L^2(\Omega))$ such that
 \[
 \int_\Omega \va _n(x)\chi_\omega(x) \int_0^T    e^{-\la_n^2 (T-r)} \tilde u(r,x) dr\, dx=\eta_n\,.
 \]

The equation
 \begin{equation}\label{equ:sect3DISTRIcontrol}
 u(r,x)-\int_0^r J(T-s,T-r) u(s,x)  ds =\tilde u(r,x)
 \end{equation}
 is a Volterra integral equation of the second kind in $L ^2(0,T;L^2(\Omega))$, hence admits a (unique) solution $u$. We conclude that if   $\eta\in L^2(\Omega)$   is reachable for the memoryless heat  equation and distributed control $\tilde u$, it  is also reachable for the equation with memory, using the control $u$ in~(\ref{equ:sect3DISTRIcontrol}).

 Hence we have approximate controllability of the memory system with distributed controls.

 The argument which proves approximate controllability with boundary control is similar, but we need some more care. Formally, the same computations as above hold with $G_n(t)$ replaced by
 \[
f_n(t)= \int_\Gamma \ga_1\va _n(x)f(t,x) d\sigma
 \]
 so that $\eta(x)$ in~(\ref{eq:defiTARGET}) is reachable by, respectively, the memoryless heat equation and by the heat equation with memory when there exists respectively $\tilde f(t,x)$ or $f(t,x)$ in $L^2(0,T;L^2(\Gamma))$ such that
\begin{align}
\label{eq:ReachETAmeroryless}
-\eta_n&= \int_\Gamma\left (\ga_1\va _n \right ) \int_0^T e^{-\la_n^2 (T-r)}   \tilde f(r ) dr\, d\sigma\,,\\
\label{eq:ReachETAmerory} -\eta_n&=\int_ \Gamma \!\!\left (\ga_1\va _n \right )  \int_0^Te^{-\la_n^2 (T-r)}  \left [
  f(r )- \!\! \int_0^r J(T-s,T-r)  f(s ) ds\right ] dr\,d\sigma\,.\!\!\!\!
\end{align}
For every fixed $n$ the integrals converge, but there is no guarantee that the series~(\ref{eq:serie ditheta}) with these coefficients will converge to an $L^2(\Omega)$-valued function which is continuous near $ T $. But, $\eta $ is by assumption reachable for the \emph{memoryless} heat equation so that the series
\[
\sum _{n=1}^{+\infty}\left [ \int_\Gamma\left (\ga_1\va _n\right ) \int_0^t e^{-\la_n^2 (t-r)}   \tilde f(r ) dr\, d\sigma\right ]\va_n(x)=w(t ,x)
\]
converges in $L^2(0,T;L^2(\Omega))$, to a function which is continuous on $(T-\varepsilon, T)$ for some $\varepsilon>0$, and $w (T,x)=\eta(x)$.

So, the same properties hold for the series of the memory system,
i.e.
\[
\sum _{n=1}^{+\infty}\left [
\int_ \Gamma\left (\ga_1\va _n \right )  \int_0^T e^{-\la_n^2 (T-r)}  \left [
  f(r )-\int_0^r J(T-s,T-r)  f(s ) ds\right ] dr\,d\sigma
\right]\va_n(x)
\]
when the  function $f$   solves the  following   Volterra integral equation in $L^2(0,T; L^2(\Gamma))$:
\[
f(r,x)-\int_0^r J(T-s,T-r)  f(s,x) ds=\tilde f(r,x)\,.
\]

This proves Theorem~\ref{Theorem1.4.}.

\section{Lack of controllability to zero}

In this section we prove that, in spite of the approximate controllability,  it is not possible to steer any initial condition $\xi$   to the smoothest possible target, i.e. $\eta=0$,    either in the distributed  or in the boundary control case.
An initial condition $\xi$ can be controlled to hit the target zero at time $T$ when there exists a (distributed or boundary) control such that for every $n$ the following equality holds:
\begin{align}
\nonumber &\int^T_0\left [ e^{-\lambda ^2_n (T-s)}-\int_0^{t-s} H_n(\tau) e^{-\lambda _n^2(T-s-\tau)}d\tau    \right ]\left (G_n(s)-f_n(s)\right )ds=\\
\nonumber&
- e^{-\lambda ^2_n T}  \xi_n+\left [
 \int_0^T e^{-\lambda _n^2(T-s)}R(s)ds  -\int_0^T e^{-\lambda_n^2(T-s)} H_n(s) ds  + \right.    \\
\label{trasfoCONTROzeroTOreach} &
\left.
  +\int R(s) \left (\int_0^{T-s} H_n(\tau) e^{-\lambda _n^2(T-s-\tau)} d\tau\right ) ds
\right ]\xi_n\,.
\end{align}
Here we intend either $F_n=0$ or $f_n=0$ for every $n$ and with the usual caveat in the boundary control case.

As we noted (see~(\ref{FormaFINALthetaN})),  using~(\ref{Eq:formadiHn}) we can write the left hand side   the same form as obtained from the memoryless heat equation:
\[ 
\int_0^T e^{-\la_n^2(T-s)}\left [
\left [G_n(s)-f_n(s)\right ]-\int_0^s J(T-\tau,T-s) \left [G_n(\tau)-f_n(\tau)\right ]d\tau
\right ] ds\,.
 \]
This fact transforms   controllability to zero of the system with memory to a suitable reachability problem \emph{for the memoryless heat equation}:  we have controllability to zero
for the system with memory  if and only if the reachable set \emph{of the memoryless heat equation} contains all elements whose Fourier coefficients are given by the right hand side of (\ref{trasfoCONTROzeroTOreach}).
Concerning these coefficients, we note the following result, which is easily proved (see~\cite{HP1}):
 
\begin{lem}
\label{Lemma4.1.} {  Let $T>0$ be such that $R(T)\neq 0$.
There exists a number $N$ such that the equations}
\begin{align*}
&- e^{-\lambda ^2_n T}  \xi_n+\left [
 \int_0^T e^{-\lambda _n^2(T-s)}R(s)ds -\int_0^T e^{-\lambda_n^2(T-s)} H_n(s) ds+ \right.  \\
 &
\left.
  +\int_0^T R(s) \left (\int_0^{T-s} H_n(\tau) e^{-\lambda _n^2(T-s-\tau)} d\tau\right ) ds
\right ]\xi_n=\frac{c_n}{\la_n^2}\,,\qquad n\geq N
\end{align*}
\emph{  are solvable for every sequence}  $\{c_n\} _{n\geq N}\in l^2([N,+\infty))$.
\end{lem}

Now we use the following important property of the memoryless heat equation:

\begin{thm} 
 Let $\tilde\omega$ be a nonempty open set  with the following property:
\begin{equation}\label{eq:condiSUomegatilde}
{\rm cl}\,\tilde \omega \subseteq \Omega\,,\quad {\rm cl}\,\tilde \omega \cap {\rm cl}\,\omega = \emptyset
\end{equation}
(the second condition is to be disregarded in the case of the boundary control, i.e. 
when $F=0$). Then, every solution $w (t,x)$ of the memoryless heat equation~(\ref{2.6}) is of class $C^\infty(( T-\varepsilon,T]\times \tilde\omega)$ (provided that  $T-\varepsilon>0$). 
\end{thm}
For completeness   we give some detail on this fact in the appendix.

Now we consider separately the case of the distributed and the boundary control.

\subsection{Lack   controllability to zero with distributed controls}

The assumptions is that $f=0$ and that the active control   is distributed in a regione $\omega\neq \Omega$.

We use Lemma~\ref{Lemma4.1.}  and we see that
if   controllability to zero holds then the targets $\eta=\theta(T  )$ which can be reached from the initial condition $\xi=0$ have the form
\[
\sum _{n\geq N} \frac{c_n}{\la_n^2} \va _n(x) + \left ( \sum _{n=1}^{N-1}\alpha_n \va _n(x)\right )\,.
\]
Here, $\{c_n\} _{n\geq N}$ is arbitrary in $l^2$ while the second term, in parenthesis, 
is related in a complicated way to the series. But, it is a linear combination of eigenfunctions, and so \emph{ it  is of class $C^\infty(\Omega)$.}

In order to prove lack of controllability to zero (for the system with memory), it is sufficient to exhibit a target $\eta(x)$ whose Fourier coefficients have the form  as the right hand side of~(\ref{trasfoCONTROzeroTOreach}), and which is not reachable by the memoryless heat equation.
Let $ \tilde\omega $ be as in~(\ref{eq:condiSUomegatilde}).
  We consider any function $\eta(x) $ with compact support in $\tilde \omega$ and  such that

\[
\eta (x)\in H^2(\tilde\omega ) \,,\qquad \eta(x)\notin H^3(\tilde\omega)\,.
\]
 Let $\eta_{\rm e}(x)$ be its extension with $0$ to $\Omega$. Then, $\eta_{\rm e}(x)\in H^2(\Omega)\cap H^1_0(\Omega)= {\rm dom}\,A$ so that, using the statement~\ref{item2:lemma:sulDomiA} in Lemma~\ref{lemma:sulDomiA},
 \begin{equation}\label{eq:expreETAperNOnullCONTR}
 \eta_{\rm e}(x) =\sum _{n=1}^{\infty}\frac{d_n}{\la_n^2}\va _n(x)=
 \sum _{n=N}^{\infty}\frac{d_n}{\la_n^2}\va _n(x)+\left ( \sum _{n=1}^{N-1}\frac{d_n}{\la_n^2}\va _n(x)\right )\,,\qquad \{d_n\}\in l^2\,.
 \end{equation}
Let us consider the sequence $ \{d_n\} _{n\geq N} $. This sequence
 can be obtained from the right hand side of~(\ref{trasfoCONTROzeroTOreach}). So,
if   controllability to zero holds for the system with memory, then there exists $u$ such that
the solution $ w $ of the memoryless system satisfies
\[
w(T,x)= \sum _{n=N}^{\infty}\frac{d_n}{\la_n^2}\va _n(x)+ \tilde w( x)\;.
\]
The function $\tilde w(x)$  is a linear combination of $ \phi_n(x) $, $ 1\leq n\leq N-1 $, hence it is of class $C^\infty(\Omega)$.

The restriction of $w(T,x)$ to $\tilde \omega $ is
\[
 \eta(x)+\left (\tilde w( x)- \sum _{n=1}^{N-1}\frac{d_n}{\la_n^2}\va _n(x)\right )\,.
\]
The sum of a function \emph{which is not} of class $H^3(\tilde\omega)$ and a function of class $C^\infty(\tilde\omega)$.

Hence $w (T,x) $ \emph{is not of class $H^3(\tilde\omega)$.} As we stated above,  this  is not possible, since   $w (T,x) \in C^\infty(\tilde\omega)$. The contradiction proves that   controllability to zero does not hold.

\subsection{Lack of  controllability to zero  under boundary controls}
The same argument can be used to prove that the target $\eta=0$ is not reachable from every initial condition,  under boundary controls. Let now $\tilde\omega$ be any subdomain such that
\[
{\rm cl}\,\tilde\omega\subseteq\Omega\,.
\] 
Every $ \eta(x)\in  \left (H^2(\Omega )\cap H^1_0(\Omega ) \right )\setminus H^3(\Omega )$ has Fourier coefficients $ \{\eta_n\} _{n\geq N}  $ of the form given by the right hand side of~(\ref{trasfoCONTROzeroTOreach}). Hence, if   controllability to zero holds there should be functions $ \eta(x)\in  \left (H^2(\Omega )\cap H^1_0(\Omega ) \right )\setminus H^3(\Omega )$ which belong to the reachable set of the memoryless system. This is not possible  since $w(t,x)\in C^\infty((T-\varepsilon,T]\times \tilde\omega)$
\emph{for every} square integrable boundary control $f$, even for those boundary control $ f $ which are not admissible, i.e. for which $w(t)\notin C((T-\epsilon,T];L^2(\Omega )) $.

\subsection*{\bf Appendix}
In this appendix, we first explain more precisely the regularity property used in the proof of the lack of  controllability to zero.  Then we report the proof of formula (\ref{Eq:formadiHn}).

\subsection*{The regularity of  $\theta(t,x)$}Let $\Omega_T= (0,T]\times\Omega$ and
\[
\mathcal{H}(\Omega_T)=\left \{ w:\Omega_T\ \mapsto \mathbb{R}\,,\quad w_t\,,\; w _{x_i x_j}\in C(\Omega_T)\right \}\,.
\]

 Let $(T,x_0)\in \Omega_T$ and let $C_\rho$ be the cylinder
 \[
 C_\rho=\left \{ (t,x)\,\ |x-x_0|<\rho\,,\quad t\in (T-\rho^2,T)\right \}\,.
 \]
 The number $\rho$ is small enough, so that $  C_{4\rho}\subseteq \Omega_T$. We denote by $\mu$ the Lebesgue measure of $C_{4\rho}$.

The following result is proved in \cite[p.~258]{DiBened}. There exist  constants $\gamma $ and $C$, which depends only on the dimension $d$ of $\Omega$, such that the following holds for every solution  $w\in \mathcal{H}(\Omega_T) $ of the heat equation
$
w'=\Delta w 
$:
 \begin{equation}
 \label{eq:condiregolPERsoluregolari}\left\{
 \begin{array}{l}
\displaystyle  \sup _{(t,x)\in C_\rho}
 |D^\alpha w|\leq \gamma \frac{C^{|\alpha|}|\alpha|!}{\rho^|\alpha| \mu}\int _{C_{4\rho}}|w(t,x)| dx\,dt\leq 
 \\ 
  \noalign{\medskip} \le \gamma \frac{C^{|\alpha|}|\alpha|!}{\rho^|\alpha|\sqrt\mu}\left [\int _{C_{4\rho}}|w(t,x)|^2 dx\,dt\right] ^{1/2}\,,\\ \noalign{\medskip}
\displaystyle  \sup _{(t,x)\in C_\rho}\left |\frac{\partial^k w}{\partial t^k}\right |\leq
  \gamma \frac{C^{2k}|\alpha|!}{\rho^|\alpha| \mu }\int _{C_{4\rho}}|w(t,x)| dx\,dt\leq  \\ \noalign{\medskip} \le \gamma \frac{C^{2k}|\alpha|!}{\rho^|\alpha|\sqrt\mu}\left [\int _{C_{4\rho}}|w(t,x)|^2 dx\,dt\right] ^{1/2}
 \end{array}\right.
\end{equation}
($\alpha$ is the multiindex of the  partial differentiation).

 Note that these inequalities can be applied to the points of $\tilde\omega$ even if a distributed control acts on $\omega$, provided that $({\rm cl}\, \tilde\omega) \cap
 ({\rm cl}\, \omega)=\emptyset$, and provided that $\partial\tilde \omega$ does not intersect $\partial\Omega$, in the case of boundary control.

 The previous inequalities have been stated for solutions which belong to $\mathcal{H}(\Omega_T)$. This is not the case if   $u\in L^2(0,T;L^2(\Omega))$ or $f\in L^2(0,T;L^2(\Gamma))$
 but, as stated in Theorem~\ref{t2.1}, every solution is the limit of a sequence of solutions which belong to $\mathcal{H}(\Omega_T)$. So, the previous inequalities can be lifted from ``smooth'' solutions to every solution given by formula (\ref{2.7}). Indeed,
 we see from~(\ref{eq:condiregolPERsoluregolari}) that $ L^2(0,T;L^2(\Omega)) $-convergence of a sequence of solutions implies  that the sequence of partial derivatives of any order is uniformly Cauchy on $ C_\rho $, thus uniformly convergent and so the partial derivatives of the limit exist. In particular, every function given by (\ref{2.7}) is of class $C^\infty(\tilde\omega)$.

\subsection*{Formula~(\ref{Eq:formadiHn})}
   
 The convolution of two functions defined on $(0,\ty)$ is
$$
f*g = \int^t_0 f(t-s) g(s) ds
$$
The convolution of $f$ with itself will be denoted as $f^{*k} = f * f^{*(k-1)}$, $f^{*1} = f$.

Let us fix an index $ n $ and let, for every $ k\geq 0 $,
 
$$
e_0 = e^{-\la_n^2 t}\,, \quad 
e_k (t) = \frac{t^k}{k!} e^{-\la_n^2 t}  .
$$
Then $e_{k+1} = e_0 * e_k$ and  $Z _n = -L * e_0$.
It is clear that
if $F$ is an integrable function and if $\tilde F = F* e_k$ then
$$
Z_n*\tilde F = e_{k+1} * (-L * F).
$$
In fact,
$$
Z_n * \tilde F = (- L * e_0) * (F * e_k) = (e_0 * e_k) * (-L * F) = e_{k+1} * (-L * F)\,.
$$

It follows that $Z_n^{*k} = (-1)^k   L^{*k} * e_{k-1}$. By \cite[p.~36]{GL},   the resolvent kernel of $Z_n$ is
 \begin{equation}\label{3.19}
\begin{array}{lcl}
H_n (t) &=& \di\sum^\ty_{k=1} (-1)^{k-1} Z_n^{*k} = - \di\sum^\ty_{k=1} L^{*k} * e_{k-1} = \\ \noa &=& -\di\int^t_0 \left(\di\sum^\ty_{k=1} L^{*k} (t-s) \di\frac{s^{k-1}}{(k-1)!}\right) e^{-\la_n^2 s} ds \end{array}
  \end{equation}
Since $|L(t)| \le M$ for $t\in [0,T]$ one has
$$
|L^{*k} (t)| \le \frac{T^k M^k}{k!}, \qu \fo t \in [0,T]
$$
so the series in (\ref{3.19}) converges uniformly in $[0,T]$. It follows from (\ref{3.19}) that
\[
\begin{array}{l}
\di\int^t_0 H_n  (t-\tau) e^{-\la_n^2  \tau} d\tau = e_0 * H  = - e_0 * \di\sum^\ty_{k=1} L^{*k} * e_{k-1} = \\ \noa = \di\sum^a_{k=1} L^{*k} * e_k = - \di\int^t_0 \left(\di\sum^\ty_{k=1} L^{*k} (t-s) \di\frac{s^k}{k!} \right) e^{-\la_n^2  s} ds   = \di\int^t_0 J(t,s) e^{-\la_n^2  s} ds  \end{array}
\]
where
\[
J(t.s)=\sum^\ty_{k=1} L^{*k} (t-s) \di\frac{s^k}{k!}
\]
is a continuous function \emph{which does not depend on $ n $.}

\end{document}